\documentclass[conference]{IEEEtran}

\usepackage{cite}
\usepackage{textcomp}
\def\BibTeX{{\rm B\kern-.05em{\sc i\kern-.025em b}\kern-.08em
    T\kern-.1667em\lower.7ex\hbox{E}\kern-.125emX}}
\usepackage{goldschmidt_ieee}
                    
\graphicspath{{figs/}} % Relative figure paths for subfiles 

\begin{document}
% ****

% Title
\title{Automatic pulse-level calibration by tracking observables using iterative learning}

\author{
    \IEEEauthorblockN{Andy J. Goldschmidt}
    \IEEEauthorblockA{\textit{Department of Computer Science} \\
    \textit{University of Chicago}\\
    Chicago, IL 60637\\
    andygold@uchicago.edu}
    \and
    \IEEEauthorblockN{Frederic T. Chong}
    \IEEEauthorblockA{\textit{Department of Computer Science} \\
    \textit{University of Chicago}\\
    Chicago, IL 60637\\
    chong@cs.uchicago.edu}
}

\maketitle
% Disallow figures in the left column on page 1
\global\csname @topnum\endcsname 0
\global\csname @botnum\endcsname 0
% \thispagestyle{plain}
% \pagestyle{plain}

% Abstract
% ========
\begin{abstract}
    Model-based quantum optimal control promises to solve a wide range of critical quantum technology problems within a single, flexible framework.
    The catch is that highly-accurate models are needed if the optimized controls are to meet the exacting demands set by quantum engineers. 
    A practical alternative is to directly calibrate control parameters by taking device data and tuning until success is achieved.
    In quantum computing, gate errors due to inaccurate models can be efficiently polished if the control is limited to a few (usually hand-designed) parameters; however, an alternative tool set is required to enable efficient calibration of the complicated waveforms potentially returned by optimal control.
    We propose an automated model-based framework for calibrating quantum optimal controls called \textit{Learning Iteratively for Feasible Tracking} (LIFT).
    LIFT achieves high-fidelity controls despite parasitic model discrepancies by precisely tracking feasible trajectories of quantum observables.
    Feasible trajectories are set by combining black-box optimal control and the \textit{bilinear dynamic mode decomposition}, a physics-informed regression framework for discovering effective Hamiltonian models directly from rollout data.
    Any remaining tracking errors are eliminated in a non-causal way by applying model-based, norm-optimal \textit{iterative learning control} to subsequent rollout data.
    We use numerical experiments of qubit gate synthesis to demonstrate how LIFT enables calibration of high-fidelity optimal control waveforms in spite of model discrepancies.
\end{abstract}


 \begin{figure}[t]
    \centering
    \includegraphics[width=\columnwidth]{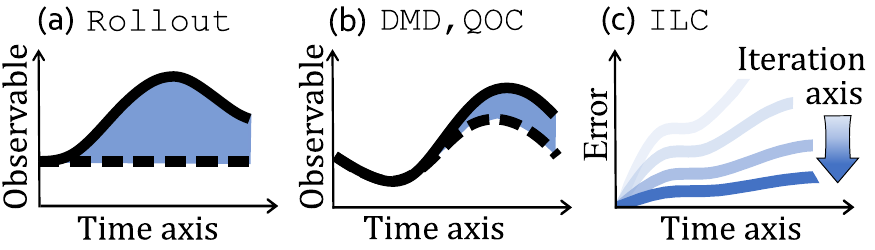}
    \caption{
    LIFT (Algorithm~\ref{alg:main}) combines the time and iteration axis through two familiar algorithms: \texttt{DMD} on the time axis and \texttt{ILC} on the iteration axis. (a)~Rollout data (solid black) from an attempt to realize a nominal reference (dotted black) are used to discover effective Hamiltonian models (\texttt{DMD}), from which (b) a feasible reference is designed via black-box quantum optimal control (\texttt{QOC}). By leveraging rollouts along the iteration axis, (c)~\texttt{ILC} is used to eliminate any remaining tracking errors.
    }
    \label{fig:ilc_main}
\end{figure}
\section{Introduction} \label{sec:introduction}
% ======
%
Quantum optimal control is a flexible, model-based optimization framework for achieving pulse-level implementations of arbitrary quantum processes. Solutions can be tailored to complicated specifications, as long as the request is formatted as an optimization problem. For example, optimal control can incorporate model detail beyond the qubit subspace, provide robustness to harmful noise and crosstalk, satisfy hardware-specific constraints or physical symmetries, discover faster gates, improve information gain in quantum metrology, and assist in device engineering and experiment design (for a recent review, see \cite{koch2022quantum}). Experimental imperfections and model uncertainties lead to suboptimal realizations of designed pulses. Instead of characterizing systems, pulse parameters can be directly calibrated using experimental data~\cite{egger2014adaptive,ferrie2015robust,sheldon2016characterizing,wittler2021integrated}. When using standard model-free calibration schemes, a hand-designed pulse with a few parameters is easier to calibrate than an optimized waveform with many. Yet, ans\"{a}tz flexibility allows optimal control to find solutions to increasingly complex tasks, so there is a tension between expressiveness and calibration.

To maximize the impact of quantum optimal control, automatic calibration frameworks are needed which are more suitable for arbitrary control waveforms. For this purpose, we propose a model-based calibration framework called LIFT (\textit{Learning Iteratively for Feasible Tracking}, Algorithm~\ref{alg:main}). Figure~\ref{fig:ilc_main} illustrates the main components of the algorithm: LIFT finds and tracks feasible reference observables along the time axis by lifting sequential rollout data onto the iteration axis. Errors from experimental imperfections and model uncertainties are eliminated in a non-causal way using sequential rollouts. For reference design, we rely on a black-box, model-based optimal controller~\cite{ball2021software,boulder2023qctrl}, and we learn drift and control Hamiltonians from rollout data via the \textit{dynamic mode decomposition} (DMD). DMD is a data-driven regression framework for discovering models directly from time-series measurement data~\cite{tu2014on}. In classical mechanics, DMD learns linear propagators for classical observables by considering time evolution in terms of Koopman-von Neumann mechanics~\cite{brunton2022modern}---the same operator-theoretic language used for quantum mechanics. For reference tracking, we implement \textit{iterative learning control} (ILC) as an optimization problem in lifted coordinates.  ILC is a widely adopted approach for realizing advanced control schemes in applications involving repetitive control under static model uncertainty~\cite{bristow2006iterative}; we take an optimization-based approach to the ILC problem~\cite{owens2005iterative,schoellig2012optimization}.

Iterative learning is a well-established part of the history of quantum optimal control~\cite{judson1992teaching}; we use the term model-based ILC to refer to hybrid optimization schemes that calibrate control solutions by combining rollout data with nominal models. For example, in quantum computing, model-based ILC has been proposed in the form of a modification to the usual gradient-based optimal control: the ILC version incorporates rollout data in the form of process tomography in order to find device-aware gradients of the terminal cost~\cite{li2017hybrid,wu2018data} (for experimental demonstrations, see~\cite{feng2018gradient,zong2021optimization}). A terminal cost is the minimum requirement for finding optimal controls for gates, but state tracking objectives have also been studied---alongside closely associated questions about reference feasibility~\cite{zhu2003quantum}. Indeed, our work is perhaps most similar to the model-based ILC implemented in~\cite{phan1999self}, where at each iteration a local input-output map was learned and used to simulate ILC offline in order to eventually track a reference observable. In our work, we use DMD to disambiguate the effect of drift and control and run ILC online in order to replicate a desired process by a proxy of tracked reference observables. Further improvements to model-based ILC and practical automatic calibration frameworks are essential for realizing the promise of optimized pulses in future quantum technologies.

% Algorithm
% =========
% Cosmetic
\RestyleAlgo{ruled}
\SetKwComment{Comment}{$\triangleright$~}{}
\newcommand\mycommfont[1]{\small\ttfamily{#1}}
\SetCommentSty{mycommfont}
\SetKwInOut{KwIn}{Input}

\begin{algorithm}[t]
\caption{\textbf{LIFT} \\Learning Iteratively for Feasible Tracking} \label{alg:main}
    \KwIn{Initial state~$\mathbf{x}(0)$, Initial model~$\mathcal{M}$, Experiment~$\mathcal{E}$, Target~$\mathbf{G}$}
    \KwResult{Calibrated control $\mathbf{u}^*$}
        \Comment{Quantum optimal control}
        {$\mathbf{x}^\text{ref.},\, \mathbf{u}_0 \gets$\textbf{QOC}$(\mathcal{M}, \mathbf{G}, \mathbf{x}(0))$}
        
        {$j \gets 0$}
        
        \While{not converged}{
            {$\mathbf{x}_j \gets $~\textbf{Rollout}$(\mathcal{E}, \mathbf{u}_j, \mathbf{x}(0))$}

            \eIf{not \normalfont{\textbf{Feasible}}$(\mathcal{M}, \mathbf{x}_j, \mathbf{u}_j)$}{
                \Comment{{\normalfont\S}\ref{sec:bidmd}: Dynamic mode decomposition}
                {$\mathcal{M} \gets$~\textbf{DMD}$(\mathbf{x}_{0:j}, \mathbf{u}_{0:j})$}
                {$\mathbf{x}^\text{ref.},\, \mathbf{u}_{j+1} \gets$~\textbf{QOC}$(\mathcal{M}, \mathbf{G}, \mathbf{x}(0))$}
            }{
                \Comment{{\normalfont\S}\ref{sec:ilc}: Iterative learning control}
                {$\mathbf{u}_{j+1} \gets$\textbf{ILC}$(\mathcal{M}, \mathbf{x}^\text{ref.}, \mathbf{x}_j, \mathbf{u}_j)$}
            }
            
            {$j \gets j+1$}
        }   
\end{algorithm}
% =============
%
\section{Methods}
% ======
%
Algorithm~\ref{alg:main} outlines our approach to calibration using LIFT (Learning Iteratively for Feasible Tracking). Our algorithm combines two well-established subroutines: $\texttt{ILC}$ implements iterative learning control as an optimization problem (Section~\ref{sec:ilc}), and $\texttt{DMD}$ implements the bilinear dynamic mode decomposition (Section~\ref{sec:bidmd}, of which the $\texttt{Feasible}$ subroutine is derivative). Section~\ref{sec:feasible} examines how tracking of reference observables can act a proxy for calibrating quantum processes. The rest of this section introduces our nominal model and describes our use of the time and iteration axis.

\subsection{Quantum control dynamics} \label{sec:qcd}
% ---------
In quantum computing, the state of an $n$-qubit system is represented by a unit vector, or ket $\ket{\psi}$, in a complex vector space representing the quantum register: $\mathcal{H} \equiv \C^2 \otimes \C^2 \otimes \cdots \otimes \C^2\cong \C^{2^n}$. Generally, an ensemble of pure quantum states can be completely characterized, in the sense of its measurement statistics, by a density matrix $\rho(t)$; that is, a non-negative self-adjoint operator in $\C^{2^n \times 2^n}$ with trace one. For example, a pure state is $\rho = \ket{\psi} \bra{\psi}$.  The controlled evolution of a density matrix can be modelled by the action of a bilinear Hamiltonian operator, $H(\mathbf{u}(t)) = H_0 + \sum_{j=1}^{J} u_j(t) H_j \in \C^{2^n \times 2^n}$, using the Liouville-von Neumann equation. Measurements can be described by a set of Hermitian observables $\{M_k\}_{k=1}^{K}$. Putting this all together, our starting point is the model
\begin{align} \label{eqn:liouville}
    &\dot{\rho}(t) = -i[H_0 + \sum_{j=1}^J u_j(t) H_j, \rho(t)] \nonumber \\
    &y_k(t) = \Tr{M_k \rho(t)}
\end{align}
We can write $\rho(t) = I / 2^n + \sum_{j=1}^{2^{2n}{-}1} \Tr(P_j \rho) P_j / 2^n$ using the traceless $n$-qubit Pauli operators, $P \in \{I, X, Y, Z\}^{\otimes n} \setminus \{I^{\otimes n}\}$. Note that the $n$-qubit Pauli operators need a prefactor to be orthonormal: $\Tr{P_j P_k} = 2^n \delta_{jk}$. They allow us to represent the density matrix using a general Bloch vector: $x_j(t) := \Tr{P_j \rho(t)}$. We can also map our dynamics onto these Pauli coordinates using a set of structure constants $[P_j, P_k] =: \sum_{\ell} 2^n \sigma_{jk\ell} P_\ell $ for the basis, such that \eqref{eqn:liouville} becomes
\begin{align} \label{eqn:vectorized}
    \dot{x}_\ell &= - i \sum\nolimits_{j,k} \sigma_{j k \ell} \Tr{P_j H} x_\ell =: \sum\nolimits_{k}[\mathbf{H}]_{\ell k} x_k \nonumber \\
    y_\ell &= \sum\nolimits_{k}\Tr{M_\ell P_k}/2^n x_k =: \sum\nolimits_{k}[\mathbf{C}]_{\ell k} x_k,
\end{align}
leading to a bilinear state space model,
\begin{align} \label{eqn:cts}
    \dot{\mathbf{x}}(t) &= (\mathbf{H}_0 + \sum_{j=1}^J u_j(t) \mathbf{H}_j)\mathbf{x}(t) \nonumber  \\
    \mathbf{y}(t) &= \mathbf{C} \mathbf{x}(t).
\end{align}
 We have $\mathbf{C}=\mathbf{I}$ when a complete readout is performed in the generalized Pauli basis.

\subsection{Discretization and linearization} \label{sec:discretize}
% ---------
The goal of Algorithm~\ref{alg:main} is to combine rollout data with a nominal model to reproduce the dynamics induced by a given quantum process. This process can be a quantum gate or other subroutine occurring over a fixed horizon spanning some $s = 0, 1, \dots,T=0{:}T$. Because we are interested in taking advantage of rollout data, we need to discretize the model in \eqref{eqn:cts} into a sequence of snapshot times at which measurements are taken. Fix a step size $\Delta t$, so $\mathbf{x}(s) := \mathbf{x}(s \Delta t)$. For the controls, enforce a zero-order hold over each step: $\mathbf{u}(t) = \mathbf{u}(s),\, \forall t \in [s \Delta t, (s+1) \Delta t)$. Now, \eqref{eqn:cts} can be integrated under a Runge-Kutta method or similar to yield a discretized nominal model,
\begin{align} \label{eqn:dst}
    \mathbf{x}(s+1) &= \mathbf{f}(\mathbf{x}(s), \mathbf{u}(s)) \nonumber \\
    \mathbf{y}(s) &= \mathbf{C} \mathbf{x}(s).
\end{align}

LIFT requires that we propose a set of observables to track, which we denote by $\mathbf{y}^\text{ref.}(0{:}T)$. This can be accomplished if an appropriate reference is known from theory, or by solving an optimal control problem using the nominal model.  Whether through insight or optimal control, our full reference corresponds to a triplet, $(\mathbf{y}^\text{ref.}(0{:}T),\, \mathbf{x}^\text{ref.}(0{:}T),\, \mathbf{u}^\text{ref.}(0{:}T{-}1))$, which is a feasible solution to \eqref{eqn:dst} according to our nominal model. The purpose of LIFT is to realize a triplet $(\mathbf{y}^\text{ref.}(0{:}T),\, \mathbf{x}^*(0{:}T),\, \mathbf{u}^*(0{:}T{-}1))$ which is a feasible solution according to the true system. 
% In Section~\ref{sec:feasible}, we discuss how feasibility connects to the ability of solutions $\mathbf{u}^*(0{:}T{-}1)$ achieving zero tracking to also realize the intended gate. For now and in Sections~\ref{sec:lift} and \ref{sec:ilc}, 
Our first goal is to establish how the nominal model is used to approximate the unknown true system for the purpose of achieving $\mathbf{u}^*(0{:}T{-}1)$. Define $\boldsymbol{\delta} \mathbf{y}(s) := \mathbf{y}(s) - \mathbf{y}^\text{ref.}(s)$, $\boldsymbol{\delta} \mathbf{x}(s) := \mathbf{x}(s) - \mathbf{x}^\text{ref.}(s)$, and $\boldsymbol{\delta} \mathbf{u}(s) := \mathbf{u}(s) - \mathbf{u}^\text{ref.}(s)$. We use our reference triplet to linearize the nominal dynamics from \eqref{eqn:dst},
\begin{align} \label{eqn:nominal}
   \boldsymbol{\delta} \mathbf{x}(s+1) &\approx 
    \mathbf{A}^\text{ref.}(s) \boldsymbol{\delta} \mathbf{x}(s) 
    + \mathbf{B}^\text{ref.}(s) \boldsymbol{\delta} \mathbf{u}(s)
    + \mathbf{r}(s) 
    \nonumber \\
    \boldsymbol{\delta} \mathbf{y}(s) &= \mathbf{C} \boldsymbol{\delta} \mathbf{x}(s).
\end{align}
where 
\begin{align}
    \mathbf{A}^\text{ref.}(s) &= \left. \partial \mathbf{f} / \partial \mathbf{x} \right|_{(\mathbf{x}^\text{ref.}(s), \mathbf{u}^\text{ref.}(s))} 
    \nonumber \\
    \mathbf{B}^\text{ref.}(s) &= \left. \partial \mathbf{f} / \partial \mathbf{u} \right|_{(\mathbf{x}^\text{ref.}(s), \mathbf{u}^\text{ref.}(s))}.
\end{align}
The feasibility of the reference snapshots is captured by the remainder $\mathbf{r}(s) := \mathbf{f}(\mathbf{x}^\text{ref.}(s), \mathbf{u}^\text{ref.}(s)) - \mathbf{x}^\text{ref.}(s+1)$, which is zero by construction.

\subsection{Lifted coordinates} \label{sec:lift}
% ---------
The key to Algorithm~\ref{alg:main} is repetition. The dynamics are assumed to be static across consecutive trials, so any tracking errors emerging from discrepancies in the nominal model will repeat in subsequent rollouts. We can obtain a useful expression of the static dynamics by expressing \eqref{eqn:nominal} using lifted coordinates---these correspond to a delay embedding with a length equal to the entire process horizon:
\begin{align}
    \boldsymbol{\delta}\mathbf{y} &= \begin{bmatrix} \boldsymbol{\delta}\mathbf{y}(0) &  \boldsymbol{\delta}\mathbf{y}(1) & \cdots &  \boldsymbol{\delta}\mathbf{y}(T) \end{bmatrix}\T \nonumber \\
    \boldsymbol{\delta}\mathbf{x} &= \begin{bmatrix} \boldsymbol{\delta}\mathbf{x}(0) &  \boldsymbol{\delta}\mathbf{x}(1) & \cdots &  \boldsymbol{\delta}\mathbf{x}(T) \end{bmatrix}\T \nonumber  \\
    \boldsymbol{\delta}\mathbf{u} &= \begin{bmatrix} \boldsymbol{\delta}\mathbf{u}(0) &  \boldsymbol{\delta}\mathbf{u}(1) & \cdots &  \boldsymbol{\delta}\mathbf{u}(T{-1}) \end{bmatrix}\T 
\end{align}
In ILC, we usually assume the ability to correctly reset the initial state, so we have $\boldsymbol{\delta}\mathbf{y}(0) = \boldsymbol{\delta}\mathbf{x}(0) = 0$. In lifted coordinates, the nominal dynamics are a static matrix of the system Markov parameters $\mathbf{C} \mathbf{A}^m \mathbf{B}$; that is,
\begin{align}
    \boldsymbol{\delta} \mathbf{x} &= \mathbf{F}^\text{ref.}  \boldsymbol{\delta} \mathbf{u} \nonumber \\
    \boldsymbol{\delta} \mathbf{y} &= \mathbf{G}  \boldsymbol{\delta} \mathbf{x} := \left(\mathbf{I}_{T{+}1} \otimes \mathbf{C}\right) \boldsymbol{\delta} \mathbf{x}
\end{align}
where
\begin{strip}
\noindent\rule{0.5\textwidth}{0.5pt}
\begin{equation}
    \mathbf{F}^\text{ref.} := \begin{bmatrix}
        0 & 0 & \cdots & 0 \\
        \mathbf{B}^\text{ref.}(0) & 0 & \cdots & 0 \\ 
        \mathbf{A}^\text{ref.}(1) \mathbf{B}^\text{ref.}(0) & \mathbf{B}^\text{ref.}(1) & \cdots & 0 \\ 
        % \mathbf{A}^\text{ref.}(2) \mathbf{A}^\text{ref.}(1) \mathbf{B}^\text{ref.}(0) & \mathbf{A}^\text{ref.}(2) \mathbf{B}^\text{ref.}(1) & \cdots & 0 \\ 
        \vdots  & \vdots & \ddots & \vdots \\ 
        \mathbf{A}^\text{ref.}(T{-}1) \cdots \mathbf{A}^\text{ref.}(1) \mathbf{B}^\text{ref.}(0) & \mathbf{A}^\text{ref.}(T{-}1) \cdots \mathbf{A}^\text{ref.}(2) \mathbf{B}^\text{ref.}(1) & \cdots & \mathbf{B}^\text{ref.}(T{-}1)
    \end{bmatrix}.
\end{equation}
\hspace*{\fill}\noindent\rule{0.5\textwidth}{0.5pt}
\end{strip}

\begin{figure*}[t]
    \centering
    \includegraphics[width=\textwidth]{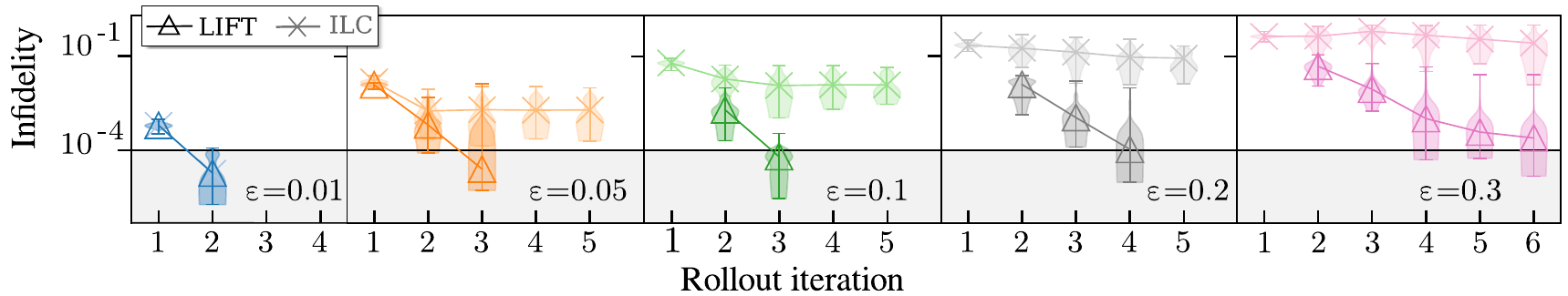}
    \caption{A violin plot comparison of gate infidelity versus number of rollouts for ILC and LIFT, where markers denote the median infidelity. Rollout iterations are used to calibrate an $X$ gate on a simulation $H(\mathbf{u}; \boldsymbol{\epsilon}_j)$ starting from a nominal model $H(\mathbf{u}; \mathbf{0})$. Subplots correspond to errors $\varepsilon$; each element in $\boldsymbol{\epsilon}_j$ is drawn from $\mathcal{N}(\varepsilon, 0.01 \varepsilon^2)$ for $300$ trials. Algorithms terminate at $99.99\%$ fidelity. A decrease in infidelity corresponds to ILC polishing, while saturation implies an infeasible reference. If a rollout was used for DMD learning in LIFT, then an infidelity was not plotted to distinguish ILC from DMD subroutines. Observe unreported infidelities at the first rollout for each $\varepsilon \ne 0.01$, demonstrating that LIFT utilizes reference redesign in these cases.}
    \label{fig:fid_vs_rollout}
\end{figure*}
\subsection{Iterative learning control} \label{sec:ilc}
% ---------
Algorithm~\ref{alg:main} considers sequential rollouts of the true dynamics---denoted in this section by $\mathbf{F}$---which are unknown and likely differ from the nominal dynamics, $\mathbf{F}^\text{ref.}$. In order to explicitly treat the use of rollouts, we append an additional iteration index $j$ to our lifted system variables, $\boldsymbol{\delta} \mathbf{y}_j:= \mathbf{y}_j - \mathbf{y}^\text{ref.}$, $\boldsymbol{\delta} \mathbf{x}_j$, and $\boldsymbol{\delta} \mathbf{u}_j$. Now, we define the quantity~$\mathbf{d}_j$ to quantify the tracking error due to discrepancies in the nominal model,
\begin{align}
    \boldsymbol{\delta} \mathbf{x}_j := \mathbf{F} \boldsymbol{\delta} \mathbf{u}_j 
    &= \mathbf{F}^\text{ref.} \boldsymbol{\delta} \mathbf{u}_j + \left(\mathbf{F} - \mathbf{F}^\text{ref.}\right) \boldsymbol{\delta} \mathbf{u}_j 
    \nonumber \\
    &=: \mathbf{F}^\text{ref.} \boldsymbol{\delta} \mathbf{u}_j  + \mathbf{d}_j
\end{align}
% NOTE: Any edge case?
Without loss of generality, $\mathbf{d}_j$ can also be assumed to include repetitive errors that we previously set to zero, like the feasibility remainder $\mathbf{r}(s)$ or initial reset $\boldsymbol{\delta}\mathbf{x}(0)$. For simplicity, in this section consider a fully observable system with $\mathbf{C} = \mathbf{I}$. The ILC subroutine in Algorithm~\ref{alg:main} is a two-step process that compensates for the current tracking error, $\mathbf{d}_j$. First, the current discrepancy $\mathbf{d}_j = \boldsymbol{\delta} \mathbf{x}_j - \mathbf{F}^\text{ref.}\boldsymbol{\delta} \mathbf{u}_j$ is obtained from the rollout data. This provides an estimate of the tracking error as a function of the control input, $\widehat{\boldsymbol{\delta} \mathbf{x}}_{j+1}(\boldsymbol{\delta} \mathbf{u}) = \mathbf{F}^\text{ref.} \boldsymbol{\delta} \mathbf{u} + \widehat{\mathbf{d}}_{j+1}$ with $\widehat{\mathbf{d}}_{j+1} = \mathbf{d}_{j}$ (a Kalman filter may also be used when $\mathbf{C} \ne \mathbf{I}$~\cite{schoellig2012optimization}).  Second, a tracking-error minimization is solved to find the optimal correction $\boldsymbol{\delta} \mathbf{u}$ to account for $\widehat{\mathbf{d}}_{j+1}$, i.e.
\begin{align} \label{eqn:ilc_opt}
    \boldsymbol{\delta} \mathbf{u}_{j+1} 
    &= \arg\min_{\boldsymbol{\delta} \mathbf{u}} \norm{ \widehat{\boldsymbol{\delta} \mathbf{x}}_{j+1} (\boldsymbol{\delta} \mathbf{u})}
    \nonumber \\
    &= \arg\min_{\boldsymbol{\delta} \mathbf{u}} \norm{ \mathbf{F}^\text{ref.} \boldsymbol{\delta} \mathbf{u} + \widehat{\mathbf{d}}_{j+1}}.
\end{align}
In practice, this convex optimization problem is solved under the state and input constraints governing the system and should include any appropriate priors like smoothness of the controls as regularization terms. For additional detail, refer to Appendix~\ref{apdx:ilc_opt}. 

The final output of LIFT is the optimal control signal, $\mathbf{u}^* = \mathbf{u}^\text{ref.} + \boldsymbol{\delta} \mathbf{u}^*$, where $\boldsymbol{\delta} \mathbf{u}_j \rightarrow \boldsymbol{\delta} \mathbf{u}^*$. This output relies on the iterative convergence of the ILC subroutine, which can be understood by considering the solution to the unconstrained convex optimization problem in \eqref{eqn:ilc_opt}. Using the lefthand Moore–Penrose inverse of the nominal dynamics, denoted $\mathbf{F}^{\text{ref.}+}$ with $\mathbf{F}^{\text{ref.}+} \mathbf{F}^\text{ref.} = \mathbf{I}$, we have
\begin{equation}
    \boldsymbol{\delta} \mathbf{u}_j
    = -\mathbf{F}^{\text{ref.}+} \mathbf{d}_j
    % = \mathbf{F}^{\text{ref.}+} \left(\mathbf{F}^\text{ref.}\boldsymbol{\delta}\mathbf{u}_j - \boldsymbol{\delta}\mathbf{x}_j \right)
    = \left( \mathbf{I} - \mathbf{F}^{\text{ref.}+} \mathbf{F}\right) \boldsymbol{\delta}\mathbf{u}_j,
\end{equation}
so that $\boldsymbol{\delta} \mathbf{u}_j$ converges to a fixed point $\boldsymbol{\delta} \mathbf{u}^*$ when the mapping $\mathbf{I} - \mathbf{F}^{\text{ref.}+} \mathbf{F}$ is contractive. This is a constraint on the closeness of our linearized nominal model and the true dynamics.

\subsection{Tracking quantum processes} \label{sec:feasible}
% ---------
In this work, we use observable tracking as a proxy for calibrating a desired quantum process. When the linearized nominal model and the true dynamics are close, the ILC subroutine converges to a triplet $\left(\mathbf{y}^\text{ref.}, \mathbf{x}^*, \mathbf{u}^* \right)$, but it does not guarantee that $\mathbf{u}^*$ replicates the quantum gate or process which induced $\mathbf{y}^\text{ref.}$. Terminal gate fidelity---a metric for the success of replication---can be estimated by measuring the final tracking error for a set of $2^n + 1$ pure states in an $n$-qubit system~\cite{reich2013minimum}. Alternatively, recall that LIFT pursues tracking along the entire time axis (as opposed to the terminal time, alone). As such, a sufficient condition for $\mathbf{u}^*$ to implement the desired gate occurs if the true optimized dynamics can be constrained to approximate the nominal reference dynamics at all times: $H^\text{true}(\mathbf{u}^*) = H^\text{nom.}(\mathbf{u}^\text{ref.})$. Practically speaking, we find that ILC is an appropriate tool for eliminating tracking errors in order to satisfy this condition, as long as (1.) the reference is \textit{feasible} on the true system, and (2.) we have more valid reference observables than controls---leading to an over-determined system and ideally enforcing $\mathbf{u}^*$ as the unique solution $H^\text{true}(\mathbf{u}^*) = H^\text{nom.}(\mathbf{u}^\text{ref.})$ at all times using less than $2^n + 1$ pure states. This first condition derives from the fact that controllable bilinear systems can involve always-on drifts which cannot be expressed in the Lie algebra generated by the controls (e.g. $H(u(t)) = Z + u(t)X$)---such models introduce an intrinsic timescale to control called the \textit{quantum speed limit}~\cite{koch2022quantum}. Feasibility demands that the quantum speed limit of the true system match that of the nominal model under the chosen time discretization. For a more complete discussion of these sufficiency criteria, see Appendix~\ref{apdx:tracking}.
% NOTE: Crosstalk is not usually in the subalgebra of control, so characterization is doing a lot of work (maybe approx., e.g. $ZZ$ in $[IY, ZX]$).

\subsection{Hamiltonian identification from snapshots} \label{sec:bidmd}
% ---------
Before utilizing the ILC subroutine (Section~\ref{sec:ilc}), we must ensure that our designed tracking reference is feasible on the true system. Therefore, we include a feasibility check in Algorithm~\ref{alg:main} comparing the nominal drift to a drift operator learned directly from the rollout data. That is, we perform Hamiltonian identification using the same data required by ILC, and---if necessary---leverage these data-driven models to redesign an improved reference. Our Hamiltonian identification is accomplished using the bilinear \textit{dynamic mode decomposition}~(DMD) subroutine. DMD takes as input a sequential snapshot record coming from experimental data or numerical simulation, formatted into a pair of offset snapshot matrices
\begin{align} \label{eqn:snapshots}
    \mathbf{X} &:=
    \begin{bmatrix}
        % | & | & & | \\
        \mathbf{x}(0) & \mathbf{x}(1) & \dots & \mathbf{x}(T{-}1) \\
        % | & | & & |
    \end{bmatrix} 
    \nonumber \\
    \mathbf{X}' &:=
    \begin{bmatrix}
        \mathbf{x}(1) & \mathbf{x}(2) & \dots & \mathbf{x}(T) \\
    \end{bmatrix},
\end{align}
and returns the optimal one-step propagator connecting the measurement sequence,
\begin{equation} \label{eqn:dmd}
    \mathbf{A}^\textrm{\footnotesize{DMD}} := \arg\min_{\tilde{\mathbf{A}}} \norm{\tilde{\mathbf{A}}\mathbf{X} - \mathbf{X}'}_F^2
\end{equation}
where $\norm{\cdot}_F$ is the Froebenius norm. Constraints can be imposed on the DMD operator using manifold optimization, as in physics-informed DMD~\cite{baddoo2023physics}---numerical differentiation can be used to replace $\mathbf{X}'$ with $\dot{\mathbf{X}}$ when necessary. In this way, skew-symmetry (or hermiticity) can be imposed, allowing the DMD subroutine to return Hamiltonian operators appropriate for \eqref{eqn:vectorized}.

Bilinear DMD is an extension of the DMD framework to bilinear quantum control systems~\cite{goldschmidt2021bilinear}. It simultaneously identifies the generators for the drift and control dynamics from rollout data under provided controls, i.e. the controls are also assembled into a snapshot matrix,
\begin{equation} \label{eqn:snapshotu}
    \boldsymbol{U} = \begin{bmatrix}
        \mathbf{u}(0) & \mathbf{u}(1) & \dots & \mathbf{u}(T{-}1) \\
    \end{bmatrix}.
\end{equation}
The bilinear DMD algorithm proceeds in the same spirit as \eqref{eqn:dmd},
\begin{equation} \label{eqn:bidmd}
    \mathbf{A}^{\scriptstyle \text{DMD}}_0,\, \mathbf{A}^{\scriptstyle \text{DMD}}_{1:\text{J}} := \arg\min_{\tilde{\mathbf{A}}} ||
    \begin{bmatrix}
        \tilde{\mathbf{A}}_0 & \tilde{\mathbf{A}}_{1:\text{J}}
    \end{bmatrix}
    \begin{bmatrix}
        \mathbf{X} \\
        \mathbf{U}{*}\mathbf{X} \\
    \end{bmatrix}
    - \mathbf{X}'||_F^2
\end{equation}
with the column-wise Khatri-Rao product, $\mathbf{U}*\mathbf{X}:=\begin{bmatrix} \mathbf{u}(1) {\otimes} \mathbf{x}(1) & \mathbf{u}(2) {\otimes} \mathbf{x}(2) &\cdots& \mathbf{u}(T) {\otimes} \mathbf{x}(T)\end{bmatrix}$. We have relied on $\mathbf{C}=\mathbf{I}$, but partial observations can also be used~\cite{otto2022learning}. The learning process can be regularized using previous nominal models~\cite{jackson2022data}.

DMD provides a data-driven approximation to the true model, but DMD can only learn operators with active controls (in addition to the always-on drift). Nonetheless, this is enough for optimal control to use DMD to design a feasible reference, at which point ILC can polish away any remaining tracking errors. Because ILC leverages a static model to sequentially correct for discrepancies, there is a trade-off in Algorithm~\ref{alg:main} between redesigning the reference trajectory using a new learned model and applying ILC to track the current reference. DMD uses a parsimonious amount of data compared to many other machine learning approaches~\cite{brunton2022modern}, and allocating one rollout (or even partial rollout) for reference redesign is often sufficient to return a useful model and feasible reference.

\begin{figure}[t]
    \centering
    \includegraphics[width=\columnwidth]{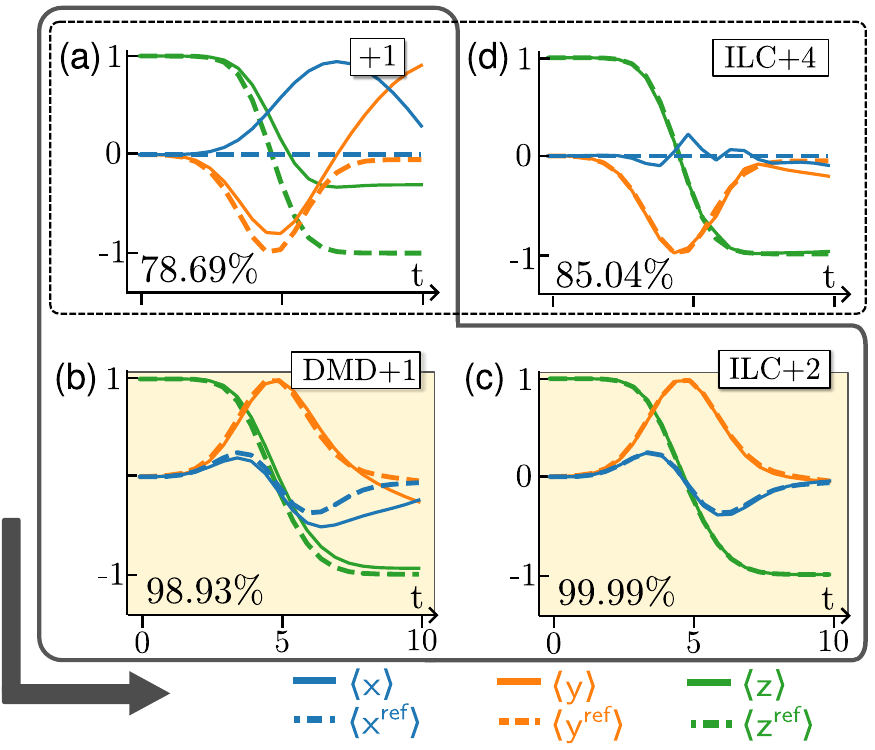}
    \caption{
    Rollout trajectories (solid colored lines) and reference trajectories (dotted colored lines) are shown at different stages of the LIFT algorithm (solid outer box, also gray triangles in Figure~\ref{fig:fid_vs_rollout}). The results are shown for a single $H(\mathbf{u}; \boldsymbol{\epsilon}_j)$, with $\boldsymbol{\epsilon}_j$ distributed about a mean of $\varepsilon=0.2$. Part~(a) shows the rollout of controls optimized using the nominal $H(\mathbf{u}; \mathbf{0})$, as well as the expected trajectory. In~(b), a redesigned feasible control---optimized using a DMD model learned from~(a)---was rolled out. Part~(c) shows successful tracking ($99.99\%$ fidelity) of the reference from (b) after ILC polishing. Part~(d) shows direct application of ILC without feasible redesign (dotted outer box, also light gray crosses in Figure~\ref{fig:fid_vs_rollout}). Stronger regularization (Appendix~\ref{apdx:ilc_opt}) removes jitter at the price of fidelity. To find the rollout iteration of each subfigure, make a running total of all annotated counts within the desired boxed routine.
    }
    \label{fig:tracking}
\end{figure}
\section{Numerical experiments} \label{sec:examples}
% ======

Consider a single qubit described by a Hamiltonian
\begin{equation}
    H(\mathbf{u}(t); \boldsymbol{\epsilon}) = \epsilon_{z} Z + u_x(t) (1 {+} \epsilon_x) X + u_y(t) (1 {+} \epsilon_y) Y
\end{equation}
using the Pauli matrices $X, Y, Z$. Take the zero error case as the nominal model, $H(\mathbf{u}; \mathbf{0})$. Suppose we design an initial reference control $\mathbf{u}^\text{ref}$ for implementing a $10$ timestep rotation gate about $X$, i.e. $R_X(\pi)=\exp{-i \pi X / 2} \cong X$ (time units are fixed by the chosen drive amplitudes). The greatest total variation under an ideal $X$ gate is found for the default initial state, $\rho(0) = \ket{0} \bra{0}$, so this is a natural choice when defining our tracking observables (Appendix~\ref{apdx:tracking}). By apply the reference control to this initial state, we induce a reference trajectory $\mathbf{x}^\text{ref}$ of the normalized Bloch vector $\mathbf{x}(t) := [\Tr{\rho(t)X}, \Tr{\rho(t)Y}, \Tr{\rho(t)Z}]\T$. In keeping with our discussion in Appendix~\ref{apdx:tracking}, we expect our three observables to be sufficient for constraining our two controls. Success is scored by the fidelity, $\mathcal{F} := |\Tr{U^\dag X}| / d$, or infidelity, $1 - \mathcal{F}$, of the resulting unitary $U := \prod_j \exp{-i H(\mathbf{u}(s_j)) \Delta t}$.

For $\boldsymbol{\epsilon} \ne 0$, a numerical simulation of $\mathbf{x}(0)$ under $H(\mathbf{u}^\text{ref}; \boldsymbol{\epsilon})$ will deviate from the intended $\mathbf{x}^\text{ref}$. The errors $\boldsymbol{\epsilon} \ne 0$ can be understood as a proxy for qubit mischaracterization and control-line distortion on a real device. To calibrate our controls and achieve our intended gate, we apply LIFT in Figure~\ref{fig:fid_vs_rollout}. We study LIFT for a range of errors: each component of a given trial $\boldsymbol{\epsilon}_j$ is drawn from $\mathcal{N}(\pm \varepsilon, \varepsilon^2 / 100)$ for $\varepsilon = 0.01, 0.05, 0.1, 0.2, 0.3$. At the price of simulated rollout data of the Bloch vector during our trajectory of duration $10$, LIFT obtains high-fidelity ($99.99\%$) gates across the full range of model discrepancies. For comparison, we also consider directly applying the ILC subroutine (Section~\ref{sec:ilc}) without including the feasibility criteria introduced in LIFT. While directly applying ILC avoids the need to spend a rollout or two on reference redesign (Section~\ref{sec:bidmd}), we observe that without this flexibility the infidelity saturates at a much higher value than the desired $99.99\%$ for all but the smallest of modeling errors. Restricting to a single trial $\boldsymbol{\epsilon}_j$ for $\varepsilon=0.2$, Figure~\ref{fig:tracking} compares the reference trajectory and rollout simulations at different stages of the full LIFT algorithm and ILC subroutine.

The total data requirements for applying LIFT can be quantified in terms of the trajectory timesteps $T$, observables $K$, and iterations $I$. The number of timesteps is fixed by the Nyquist frequency of the tracking reference, the number of observables is set by their ability to constrain the present controls (Appendix~\ref{apdx:tracking}), and the number of iterations are related to the DMD accuracy and the desired tracking error~\cite{lu2020prediction}.

\section{Conclusion} \label{sec:future}
% ======
In this work, we introduce LIFT by combining the time and iteration axis through two familiar algorithms: ILC on the iteration axis and DMD on the time axis. LIFT realizes pulse calibration by using trajectory tracking as a proxy for replicating desired quantum processes. Ideal future applications of LIFT calibration on devices will involve situations where the complex pulses from optimal control offer compelling improvements over hand-design~\cite{shi2019optimized}.
% TODO: Quantify /when/ this will happen?
The key to practical implementation of LIFT will be managing data. Toward this end, methods for managing observables when $\mathbf{C} \ne \mathbf{I}$ are essential~\cite{schoellig2012optimization,otto2022learning}. By leveraging additional physical knowledge and more precise Hamiltonian learning, we can focus on expressly learning the uncontrolled drift. Collaborative learning of control and tomography may also lower measurement requirements~\cite{ding2021collaborative}. Combining LIFT with model predictive control~\cite{zhu2003quantum,goldschmidt2022model} may allow for added flexibility under reference infeasibility. ILC approaches like data-driven gradient descent~\cite{wu2018data} can use rollouts to approximate trajectory costs in addition to the terminal cost~\cite{vemula2022effectiveness}, and adapting hybrid quantum-classical approaches to these settings may be useful~\cite{li2017hybrid}.  Adding other learning frameworks to the ILC subroutine in LIFT may improve scalability~\cite{abbeel2006using,agarwal2021regret}. 

\section*{Acknowledgments} \label{sec:ackn}
% ======
AJG thanks Aaron Trowbridge and Zac Manchester for helpful discussions on iterative learning. 
This work is funded in part by EPiQC, an NSF Expedition in Computing, under award CCF-1730449; in part by STAQ under award NSF Phy-1818914; in part by the US Department of Energy Office of Advanced Scientific Computing Research, Accelerated Research for Quantum Computing Program; in part by the NSF Quantum Leap Challenge Institute for Hybrid Quantum Architectures and Networks (NSF Award 2016136); and in part based upon work supported by the U.S. Department of Energy, Office of Science, National Quantum Information Science Research Centers.
FTC is Chief Scientist for Quantum Software at Infleqtion and an advisor to Quantum Circuits.

% Appendices
% **********
\appendices
% \raggedbottom\sloppy

\section{Iterative learning control: Optimization} \label{apdx:ilc_opt}
% ======
Equation~\ref{eqn:ilc_opt} can be written as a general quadratic program~\cite{schoellig2012optimization}
\begin{align} \label{eqn:ilc_opt_full}
    \boldsymbol{\delta} \mathbf{u}_{j+1} 
    &= \arg\min_{\boldsymbol{\delta} \mathbf{u}}
    ||\mathbf{W} ( \mathbf{F}^\text{ref.} \boldsymbol{\delta} \mathbf{u} + \widehat{\mathbf{d}}_{j+1} )||_2
    + \lambda ||\mathbf{D} \boldsymbol{\delta} \mathbf{u}||_2
    \nonumber \\
    &\,\quad \text{s.t.} \quad \abs{\mathbf{u}^\text{ref.} + \boldsymbol{\delta} \mathbf{u}} \le \mathbf{u}_\text{sat.}, 
    \quad \abs{\boldsymbol{\delta} \mathbf{u}} \le \boldsymbol{\delta} \mathbf{u}_\text{sat.},
\end{align}
where several new design choices have been introduced. The prefactor $\mathbf{W}$ allows us to assign different weights to certain times or observables. The penalty term (regularized by $\lambda$) allows us to ensure smoothness by setting $\mathbf{D}$ as a numerical derivative. The saturation constraints---$\mathbf{u}_\text{sat.}$, $\boldsymbol{\delta} \mathbf{u}_\text{sat.}$---enforce hardware limitations. Additional generalizations are possible, e.g. expanding $\mathbf{x}$ and $\mathbf{u}$ into a function basis.

\section{Replicating gates by tracking observables} \label{apdx:tracking}
% ======
Ideally, LIFT (Algorithm~\ref{alg:main}) realizes a quantum gate. Practically, LIFT converges to a triplet $\left(\mathbf{y}^\text{ref.}, \mathbf{x}^*, \mathbf{u}^* \right)$ that is feasible on the true system by tracking the triplet $\left(\mathbf{y}^\text{ref.}, \mathbf{x}^\text{ref.}, \mathbf{u}^\text{ref.} \right)$ which is feasible on the nominal model. A sufficient condition for $\mathbf{u}^*$ to realize the desired gate occurs if the true dynamics approximate the nominal dynamics at all times: $H^\text{true}(\mathbf{u}^*)=H^\text{nom.}(\mathbf{u}^\text{ref.})$. To study this claim, consider our nominal model~\eqref{eqn:liouville}, projected onto the generalized Pauli basis introduced in~\eqref{eqn:vectorized}, with $H_j = \sum\nolimits_i \Tr{P_i H_j} P_i / 2^n =: \sum\nolimits_i h_{ij} P_i / 2^n$. Hence, the nominal model $H^\text{nom.}(\mathbf{u}^\text{ref.})$ is
\begin{equation}
    H(\mathbf{u}^\text{ref.}; \mathbf{0})
    = \sum_{i=1}^{4^n-1} \left(h_{i0} {+} \sum\nolimits_j u^\text{ref.}_j h_{ij} \right) P_i/2^n
\end{equation} 
while $H^\text{true}(\mathbf{u}^{*})$ carries model discrepancies $\epsilon_{ij}$,
\begin{equation}  \label{eqn:ham_diff}
    H(\mathbf{u}^*; \mathbf{E}) := \sum_{i=1}^{4^n-1} \left((1 {+} \epsilon_{i0}) h_{i0} {+} \sum\nolimits_j u^*_j (1 {+} \epsilon_{ij}) h_{ij} \right) P_i/2^n.
\end{equation}
This framework approximates common coherent modeling errors like qubit frequency drift, crosstalk, and control line distortions. Define $\mathcal{I} := \{i: 1 \le i \le 4^n-1,\, h_{ij} \ne 0\}$ as the set of controllable Pauli axis and $\bar{\mathcal{I}}$ as the complement, in order to distinguish pure drift from control.  For the nominal Hamiltonian to match the true Hamiltonian, we require
\begin{equation} \label{eqn:match_cond}
    H(\mathbf{u}^*; \mathbf{{E}}) \overset{!}{=} H(\mathbf{u}^\text{ref.}; \mathbf{0})
    \Leftrightarrow \sum_{i' \in \bar{\mathcal{I}}} \alpha_{i'} P_{i'} + \sum_{i \in \mathcal{I}} \beta_i P_i \overset{!}{=} 0
\end{equation}
where $\alpha_{i'}:=\epsilon_{i'0} h_{i'0}$ and, recalling $\mathbf{u}^* = \mathbf{u}^\text{ref.} + \boldsymbol{\delta} \mathbf{u}^*$,
\begin{align} \label{eqn:ham_suff}
    [\boldsymbol{\beta}]_i &:= [\mathbf{B}]_i \boldsymbol{\delta} \mathbf{u}^* + [\mathbf{b}]_i 
    \nonumber \\
    &:= \sum\nolimits_j (1 + \epsilon_{ij}) h_{ij} \delta u^*_j + \left(\epsilon_{i0} h_{i0} + \sum\nolimits_j \epsilon_{ij} u^\text{ref.}_j \right).
\end{align}
For \eqref{eqn:match_cond} to hold, it is sufficient for $\boldsymbol{\alpha} = 0$ and $\boldsymbol{\beta} = 0$. A reasonable model for quantum devices assumes each control is attached to a linearly independent---often even unique---Pauli axis (i.e. $h_{ij} = \delta_{ij} h_{ij}$), so we have a matrix $\mathbf{B} \in \R^{|\mathcal{I}| \times J}$ invertible by construction. Therefore, there exists a solution $\boldsymbol{\delta} \mathbf{u}^*$ for which $\boldsymbol{\beta} = 0$.

What remains is to uncover criteria which enforces these sufficient conditions when we successfully attain zero tracking error. Starting from~\eqref{eqn:liouville}, we now presuppose successful tracking of a given set of initial states, $\{\rho_\ell(0)\}_{\ell=1}^L$. 
% NOTE: Technically, you only guarantee $\mathf{y}^\text{ref.}$ tracking, but we will set C=I.
Under this assumption, $\rho^*_\ell(t) = \rho^\text{ref.}_\ell(t)$ and $\dot{\rho}^*_\ell(t) = \dot{\rho}^\text{ref.}_\ell(t)$, such that for observables $\{M_k\}_{k=1}^K$,
\begin{equation} \label{eqn:track}
    \Tr{M_k \left[H(\mathbf{u}^*(t); \mathbf{E}) - H(\mathbf{u}^\text{ref.}(t); \mathbf{0}), \rho^\text{ref.}_\ell(t) \right]} = 0.
\end{equation}
Define
\begin{align} \label{eqn:structure}
    [\mathbf{S}]_{k\ell,i} := \Tr{M_k [P_i, \rho^\text{ref.}_\ell(t)]}
    % = \sum_{j,m} x^\text{ref.}_{\ell j} \sigma_{ijm} \Tr{M_k P_m} / 2^n
    = \sum_{j,m} x^\text{ref.}_{\ell j} \sigma_{ijm} [\mathbf{C}]_{km}
\end{align}
for $i \in \mathcal{I}$, so $\mathbf{S} \in \R^{K L \cross |\mathcal{I}|}$ for $K$ observables, $L$ initial states, and $|\mathcal{I}|$ control axes, respectively. Equivalently, let $[\bar{\mathbf{S}}]_{k\ell,i'}$ on the set $\bar{\mathcal{I}}$ such that $\bar{\mathbf{S}}\boldsymbol{\alpha}$ captures the pure drift contribution, so \eqref{eqn:track} can be written
\begin{equation}
    \bar{\mathbf{S}}\boldsymbol{\alpha} + \mathbf{S} \boldsymbol{\beta} = 0.
\end{equation}
We define a trajectory to be \textit{feasible} on the true system if the uncontrollable drift operators match, i.e. $\boldsymbol{\alpha} = 0$. If feasible, only $\boldsymbol{\beta}=0$ must be enforced for \eqref{eqn:match_cond} to hold, and $\mathbf{S} \boldsymbol{\beta}=0$ is a system of $K \cdot L$ linear equations in $|\mathcal{I}|$ unknowns (usually, $|\mathcal{I}|=J$, the number of controls).  Therefore, we can construct $\mathbf{S} \boldsymbol{\beta} = 0$ as an over-determined system by choice of tracking observables and initial states, making nonzero solutions unlikely. Explicitly, each row in $[\mathbf{S}]_{k\ell,i}$ contributes one linearly independent constraint determined by $\sigma_{ijk}$ with values set by the reference snapshots, $x^\text{ref.}_{\ell j}$.
% Careful---proof might be needed.
As long as our desired gate sufficiently excites $x^\text{ref.}_{\ell j}$, our equation coefficients remain nonzero and distinct (good initial tracking states should maximize their total variation under the reference dynamics). Last, note that $\boldsymbol{\beta}(t)$ must be continuous---a constraint incorporated into our ILC optimization in Appendix~\ref{apdx:ilc_opt}---further reducing the possibility of finding nonzero solutions.

% Bibliography
% ************
\newpage
\bibliographystyle{IEEEtran}
\bibliography{universal}
% ===========

\end{document}